# Characteristic lengthscales of the electrically-induced insulator-to-metal transition

Theodor Luibrand[†1], Adrien Bercher[†2], Rodolfo Rocco[†3], Farnaz Tahouni-Bonab[1], Lucia Varbaro[2], Carl Willem Rischau[2], Claribel Domínguez[2], Yixi Zhou[2], Weiwei Luo[2], Soumen Bag[3], Lorenzo Fratino[3,4], Reinhold Kleiner[1], Stefano Gariglio[2], Dieter Koelle[1], Jean-Marc Triscone[2], Marcelo J. Rozenberg[3], Alexey B. Kuzmenko[2], Stefan Guénon[1] and Javier del Valle[*2,5]

[1]Physikalisches Institut, Center for Quantum Science (CQ) and LISA+, Eberhard Karls Universität Tübingen, Auf der Morgenstelle 14, Tübingen 72076, Germany
[2]Department of Quantum Matter Physics, University of Geneva, 24 Quai Ernest-Ansermet, 1211 Geneva, Switzerland
[3]Laboratoire de Physique des Solides, UMR8502 CNRS - Université Paris-Sud, Université Paris-Saclay, 91405 Orsay Cedex, France
[4]Laboratoire de Physique Théorique et Modélisation, CNRS UMR 8089, CY Cergy Paris Université, 95302 Cergy-Pontoise Cedex, France
[5]Department of Physics, University of Oviedo, C/ Federico García Lorca 18, 33007 Oviedo, Spain

[†]These authors contributed equally to this work

*Corresponding author: javier.delvalle@unige.ch

**Abstract**

Some correlated materials display an insulator-to-metal transition as the temperature is increased. In most cases this transition can also be induced electrically, resulting in volatile resistive switching due to the formation of a conducting filament. While this phenomenon has attracted much attention due to potential applications, many fundamental questions remain unaddressed. One of them is its characteristic lengths: what sets the size of these filaments, and how does this impact resistive switching properties. Here we use a combination of wide-field and scattering-type scanning near-field optical microscopies to characterize filament formation in $NdNiO_3$ and $SmNiO_3$ thin films. We find a clear trend: smaller filaments increase the current density, yielding sharper switching and a larger resistive drop. With the aid of numerical simulations, we discuss the parameters controlling the filament width and, hence, the switching properties.

**Introduction**

Many correlated materials, such as the vanadate and rare-earth nickelate families, are well-known for their insulator-to-metal transition (IMT) [1–3]. The transition into the metallic state can be induced by increasing temperature, adding dopants or applying high pressures [4–6], but it can also be triggered electrically [7–11]. A large enough applied voltage or current can create a percolative metallic filament due to Joule heating [12], drastically reducing the resistance of the system [13–19]. We must note that this filament is not caused by the diffusion of ions under strong electric fields as commonly observed in resistive RAMs (random access memories) [20], but rather by a local phase transition from insulator to metal. When the voltage (current) is removed, the filament disappears, resulting in volatile resistive switching [21]. This phenomenon has attracted a lot of attention due to promising applications in emerging technologies, such as emulating neuronal spiking for neuromorphic computing [22–28], probabilistic bits for stochastic computing [29–31], or serving as electrooptical switches for optoelectronics [32–36]. In spite of this, many fundamental aspects of the electrically induced IMT are poorly understood. One of the most salient issues is the typical lengthscale of this process: what sets the size of the metallic filaments, or the number of them? And similarly, how do these characteristic lengths affect the resistive switching properties i.e. the sharpness of the switch ($\partial V/\partial I$) or the total resistive drop? Understanding this is not only of fundamental interest, but also key for designing device applications.

Here, we use a combination of wide-field optical microscopy [37] and scattering-type scanning near-field optical microscopy (s-SNOM) [38] to characterize filament lengthscales during the electrically-induced IMT in $NdNiO_3$ and $SmNiO_3$ microdevices. These compounds are two well-known members of the rare earth nickelate family [2,6]. They both display an IMT concomitant with a structural phase transition, but there are rather important differences between the two. $NdNiO_3$ has a sharp IMT around 120 K (depending on epitaxial strain) with a resistivity drop of more than two orders of magnitude (Fig. 1a). $SmNiO_3$ on the other hand, displays a smooth IMT around 400 K, with a one order of magnitude resistivity change [2,6]. Such different IMTs allow us to contrast the results from both materials and to determine which parameters govern filament lengthscales.

**Methods**

We fabricated two-terminal microdevices on top of our $NdNiO_3$ and $SmNiO_3$ films, as depicted in Fig. 1b. The nickelate films (~40 nm thick, Fig. S1) were grown on $LaAlO_3$ (001) substrates using off-axis magnetron sputtering. We used a combination of optical lithography and ion etching to define isolated 360 μm x 130 μm nickelate islands, on top of which we patterned two planar Pt electrodes using a second optical lithography and on-axis Pt sputtering. The electrodes are 20 μm wide, with a 20 μm gap between them (10 μm x 10 μm for the s-SNOM experiments). The IMT can be triggered by applying a large enough voltage or current across the gap. To image this phenomenon, we take advantage of the large reflectivity change across the IMT [10]. We use optical microscopy to capture the distribution of metallic/insulating domains in the gap between electrodes [17]. We do so *in operando* i.e. while applying a variable bias current, which allows us to capture clear images of the percolating filament. More details about the fabrication process, device characterization, and the experimental methods can be found in the supplementary information.

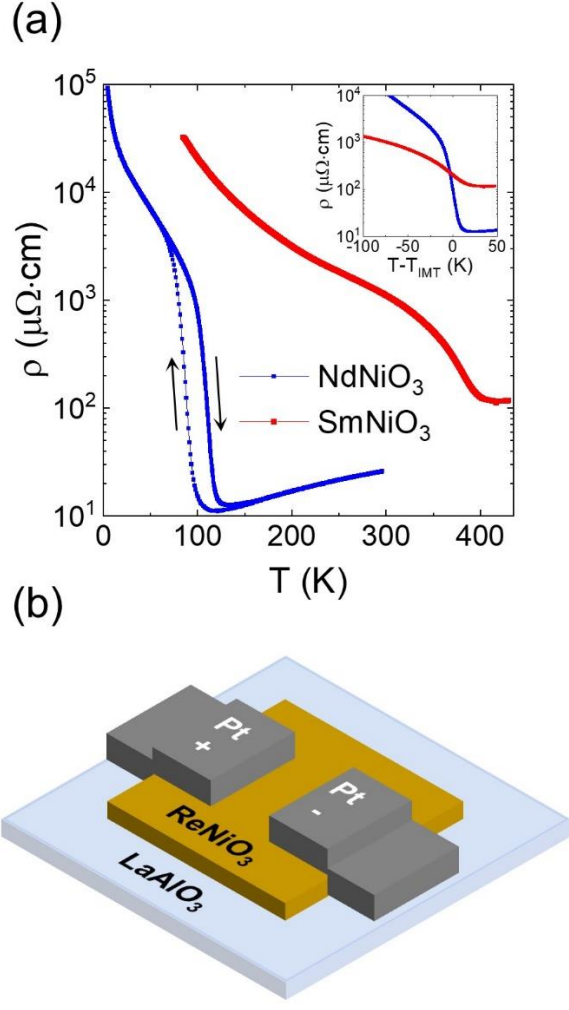

FIG. 1. Sample characteristics. (a) Resistivity vs temperature for ∼40 nm thick NdNiO$_3$ (blue) and SmNiO$_3$ (red) thin films. Inset: Resistivity plotted as a function of $T$-$T_{IMT}$. $T_{IMT}$ was calculated finding the maximum of $\partial \log(\rho)/\partial T$, where $\rho$ is the resistivity. Only the warm up branch is shown. (b) Schematic representation of the two-terminal devices. Nickelate islands (brown) were patterned on top of a LaAlO$_3$ substrate (blue). Two platinum electrodes (grey) were used to electrically trigger the IMT. The schematic is not at scale.

**Experimental results**

Fig. 2a shows the voltage $V$ as a function of the current $I$ in a NdNiO$_3$ microdevice at several temperatures below the IMT temperature i.e. the film is in the insulating state when no current is applied. As the current is ramped up, the voltage rises steeply until a threshold is reached, after which a steep voltage reduction takes place. Such drop marks the moment when the electrically-induced IMT occurs and a filament percolates between the electrodes. This is a well-known phenomenon [13–19], and it can be readily observed with our imaging set-up. The filament widens when current is further increased, shrinks for decreasing current and disappears at low enough current values (Fig. S2 and supplementary videos), in accordance with the volatile $V$-$I$ curves in Fig. 2a. We must note that some of the $V$-$I$ curves feature two discontinuities. In the current range between them, the system is not stationary but rapidly oscillates between a high and a low resistance state, as further discussed in section 3 of the supplementary information. As expected, we find the filament width to be strongly dependent on the bias current, similarly to what has been reported in previous works [14,15,17]. The bulk of this paper focuses on how other factors, such as temperature or material properties, also play a key role at setting filament size.

Fig. 2a shows a clear trend: the voltage drop becomes sharper and larger as the temperature is lowered. This feature is also observed in SmNiO$_3$ microdevices (Fig. 2d), where resistive switching at room temperature is modest, very gradual and without discontinuities. As the temperature is lowered, the drop becomes steeper and discontinuous. Comparing NdNiO$_3$ and SmNiO$_3$, it is clear that resistive switching is sharper for the former. Figs. 2b and 2e show optical microscopy images of filaments in NdNiO$_3$ and SmNiO$_3$, respectively, for the same applied current, $I = 20$ mA. Images at two different temperatures are displayed for each material, showing distinctively thinner filaments at lower temperatures in both cases. This is better appreciated in Figs. 2c and 2f, where filament width (for $I = 20$ mA) is plotted as a function of temperature. Lower temperatures yield thinner filaments and, therefore, higher current densities. Moreover, comparing NdNiO$_3$ and SmNiO$_3$, we can see that filaments are much thinner in NdNiO$_3$. Therefore, for a fixed bias current, the filament size is strongly dependent on the material and the base temperature. Fig. 2 as a whole establishes a strong connection between filament size and $V$-$I$ characteristics: thinner filaments (higher current densities) lead to sharper and larger resistive switching.

While optical microscopy is a versatile tool to visualize metallic/insulating areas at multiple temperatures and currents, its spatial resolution is limited by diffraction. In order to get more detailed images, we performed *in-operando* cryogenic s-SNOM measurements in NdNiO$_3$ devices, using a set-up such as the one depicted in Fig. 3a. The spatial resolution of this atomic force microscopy (AFM)-based technique is limited only by the tip radius (~20 nm) [38], allowing us to obtain high-resolution AFM and near-field images of the filaments. Figs. 3b and 3c show topography and SNOM images at 18 K and 70 K, respectively. Images for 0, 10 and 20 mA are displayed. For similar currents, filaments are thinner at lower temperatures, in accordance to wide-field optical images in Fig. 2. Moreover, s-SNOM allows us to resolve clear qualitative differences between both temperatures. Images taken at 18 K show a single, intense filament percolating between electrodes, while at higher temperature multiple filaments appear. Thus, lower temperatures favor a *winner-takes-all* situation in which a single filament carries all the current.

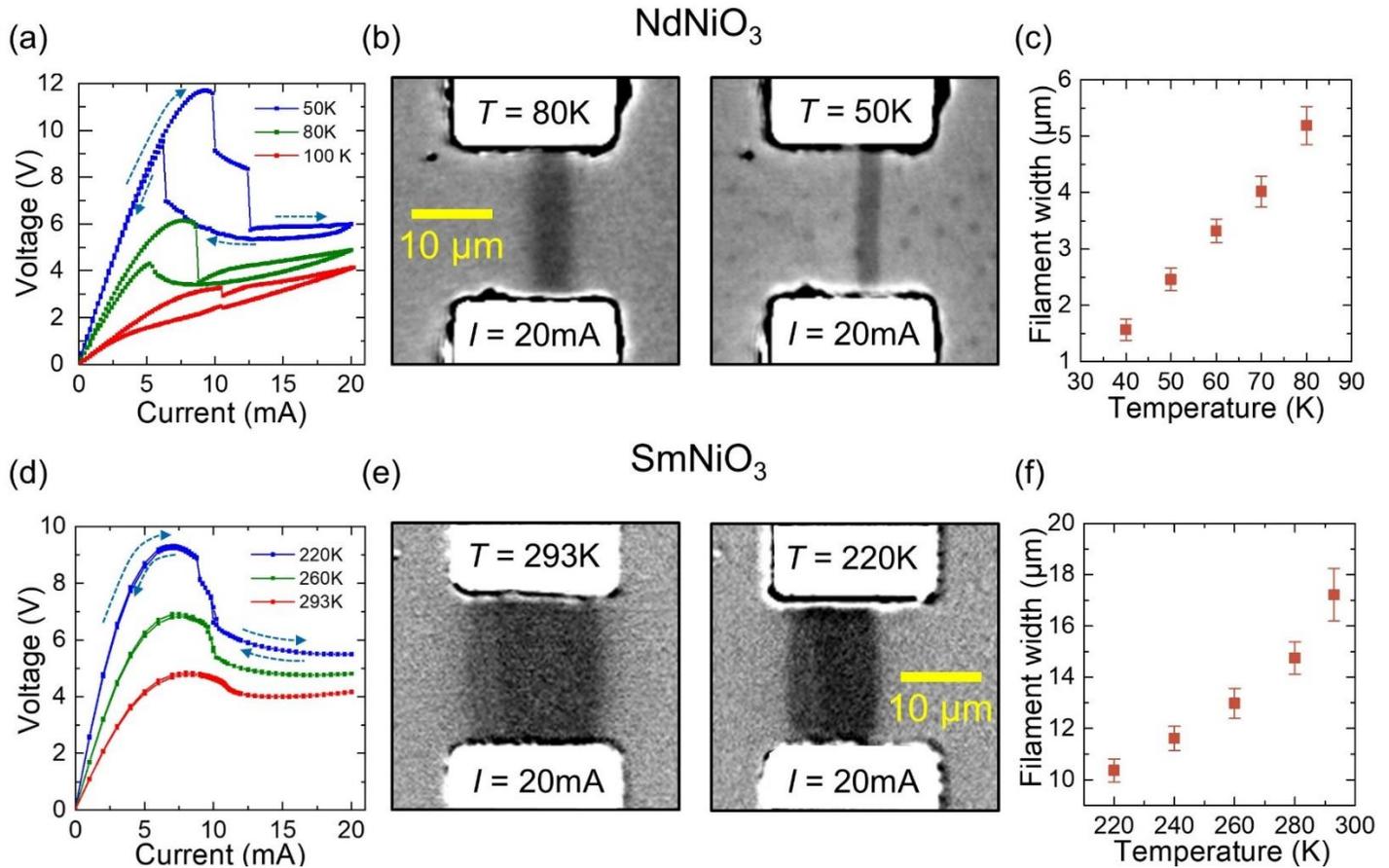

FIG. 2. Connection between resistive switching properties and filament size. (a), (d) Voltage vs current curves for NdNiO$_3$ and SmNiO$_3$ microdevices, respectively. Several temperatures are plotted for each material. (b), (e), Wide-field optical microscopy images of filaments in NdNiO$_3$ and SmNiO$_3$, respectively. Current is 20 mA for all four images. Two different temperatures are shown for each material. For the NdNiO$_3$ images, reflectivity was normalized using a region far from the gap. For SmNiO$_3$, differential images are shown, where at each point, the reflectivity at $I$=0 mA is subtracted. (c), (f), Filament width vs temperature at $I$=20 mA. The width was determined using a gaussian fitting of linescans perpendicular to the filament direction, taking the full-width-half-maximum as filament width. The error bars show the standard deviation of the distribution of widths.

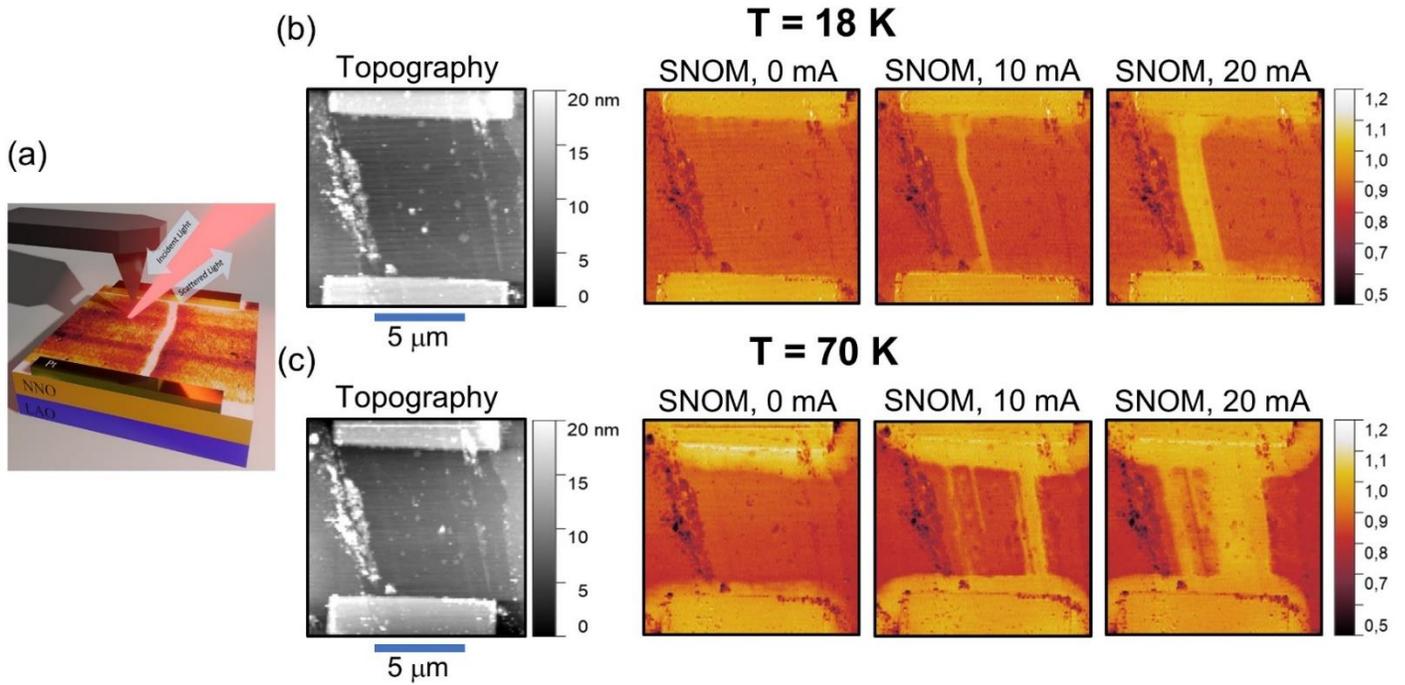

FIG. 3. High resolution s-SNOM imaging and presence of multiple percolating filaments. (a) Schematic representation of the s-SNOM set-up. Infrared light (wavelength 10 μm) is focused at a metal-coated AFM tip, which further focuses light into an area comparable to the tip radius (~20 nm). The tip-scattered signal is determined by the optical conductivity of the area of the material directly underneath the tip, allowing to get high resolution images of metallic (high signal) and insulating (low signal) domains. The AFM is working in tapping mode and the s-SNOM signal is detected at the 3rd order harmonic to filter out the far-field component as described in the supplementary. (b), (c) Topography and s-SNOM amplitude images for $NdNiO_3$ microdevices at $T$=18 K and $T$=70 K, respectively. The s-SNOM signal was normalized using the Pt electrode as reference. S-SNOM data for three different current values is shown.

**Resistor network simulations**

To understand these results, we performed numerical simulations in which we model our system as a two-dimensional resistor network (Fig. 4a) [10,24]. Each node in the network can be either metallic or insulating, depending on the local temperature and a Landau free energy functional that mimics the IMT. The insulating state resistivity increases as the temperature is decreased, following a variable range hopping dependence (Inset Fig. 4b). Currents and voltages at each node are calculated by solving Kirchhoff's laws. Local temperature is updated in each simulation time step, considering Joule heating and heat conduction. A more detailed description of the simulations can be found in the methods section of the supplementary material. This simple model reproduces experimental results with only a few parameters (Fig. 4b), allowing us to identify which ones play a key role.

Fig. 4c shows simulated 2-dimensional resistivity maps of the devices for three different values of current and base temperature. As expected, filaments are strongly dependent on the bias current. Also, similarly to the experiments, filaments become narrower as the base temperature is lowered. The material confines current flow into a smaller region at lower temperatures. This in turn induces higher current densities, increasing local Joule heating and greatly affecting the temperature distribution across the device, as can be seen in Fig. 4d. As the filament narrows, its inner temperature increases. Therefore, thinner filaments induce a stronger current-temperature feedback, which is a key factor controlling switching dynamics [10,18,29]. A strong feedback makes the device susceptible to runaway effects which manifest as discontinuities in the experimental *V-I* curves [29]. As a result, the thinner the filament and the higher its current-temperature feedback, the larger the discontinuities in the *V-I* curves. Experimentally, such discontinuities are observed to be bigger and more frequent at lower temperatures, and especially for $NdNiO_3$. These are the same conditions in which the thinnest filaments are observed.

**Discussion**

Since filament size determines switching properties, it is key to identify which parameters control the material's ability to confine current into smaller or larger areas. Here we analyze two contributions: the resistivity difference across the IMT and the thermal conductivity of the substrate. The resistivity contrast between the insulating ($\rho_{ins}$) and metallic ($\rho_{met}$) phases is expected to play a major role, since it corresponds to the resistivities outside and inside the filament. When $\rho_{ins} \gg \rho_{met}$, the current is strongly focused into the filament, reducing Joule heating outside. This keeps the insulating areas cold and confines the filament in a small region. But as $\rho_{ins}$ decreases, the insulator becomes leaky, allowing current flow and power dissipation outside the filament and reducing its confinement.

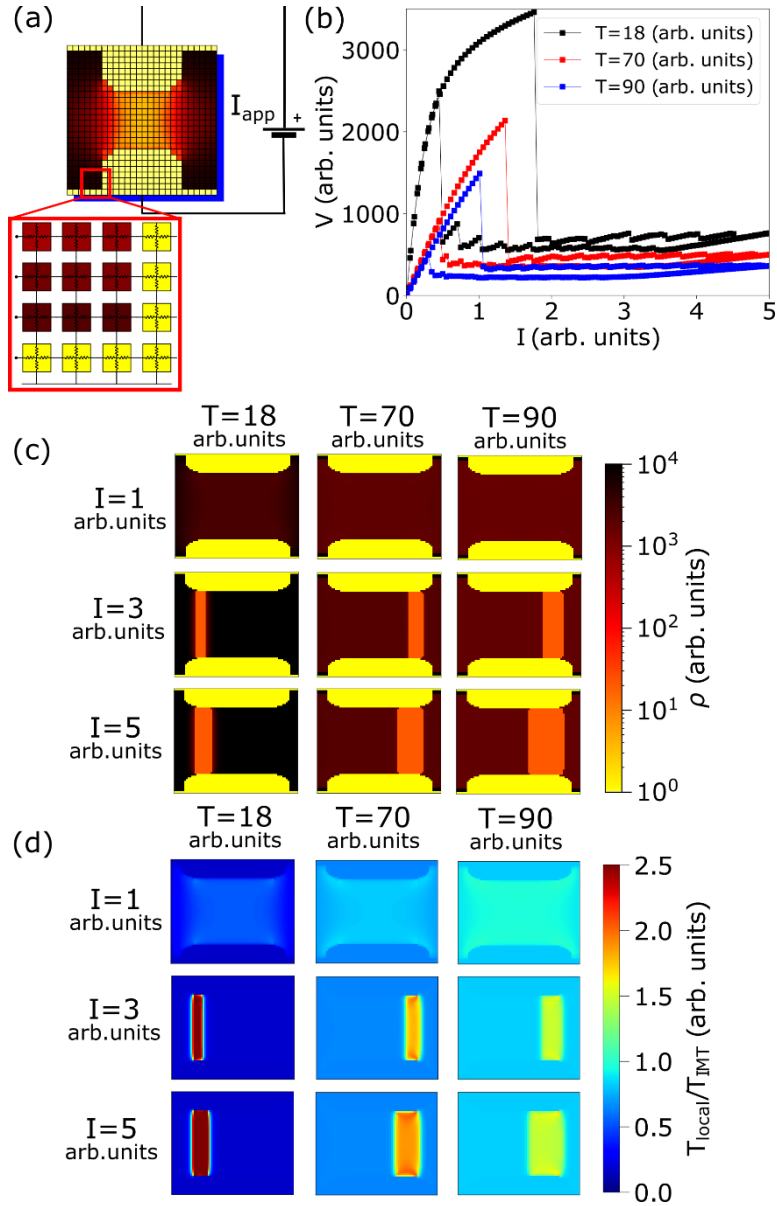

FIG. 4. Resistor network simulations and current focusing effect. (a) Schematic representation of the simulated resistor network (size *W* x *L*). Low resistance electrodes (yellow) define an oxide gap where individual nodes could be either metallic (orange) or insulating (dark brown), as described in the methods section of the supplementary material. (b) Simulated voltage vs. current curves for three different temperatures: 18 a.u. (black), 70 a.u. (red) and 90 a.u. (blue). Inset: Simulated resistance vs. temperature of the device. (c) Simulated, two-dimensional resistivity plots for all combinations of three currents (1, 3 and 5 a.u.), and three device base temperatures (18, 70 and 90 a.u.). Resistivity is plotted in logarithmic colorscale. (d) Simulated, two-dimensional temperature plots for all combinations of three currents (1, 3 and 5 a.u.), and three device base temperatures (18, 70 and 90 a.u.). Temperature is plotted in linear colorscale and normalized to the transition temperature (120 a.u.).

The temperature dependence of the resistivity implies that as temperature increases, so does power dissipation outside the filament, increasing its width. For instance, the voltage at $I$=5 a.u. is half as large for $T$=90 a.u. than for $T$=18 a.u. (Fig. 4b), but the insulating resistivity is nearly 8 times smaller. As a result, power dissipation outside the filament is increased by a factor of 2 at $T$=90 a.u. vs $T$=18 a.u. These differences in resistivity explain not only the temperature dependence of the filament size, but also the differences observed between NdNiO$_3$ and SmNiO$_3$. The former has a larger resistivity change across the IMT (Fig. 1a) [6], and is therefore expected to focus current into thinner filaments. Furthermore, the presence of single or multiple percolating filaments (Fig. 3) can be understood in a similar light. For $\rho_{ins} \gg \rho_{met}$, the current will crowd through the first hotspot that metallizes, favoring a *winner-takes-all* scenario.

But the resistivity drop across the IMT is not the only mechanism that can explain the differences in filament size. Thermal properties are also expected to play a crucial role, and they can provide a similarly satisfactory explanation. The thermal conductivity of LaAlO$_3$ is not constant, but decreases from 0.6-0.7 W/cm·K at 60 K to ~0.15 W/cm·K at 300 K [39]. This means that the substrate can more efficiently evacuate heat at lower temperatures, keeping cold the areas surrounding the filament. The filament is therefore confined to a smaller region, as simulations have recently shown [18]. Therefore, thermal properties can also explain the smaller filament size at lower temperatures, as well as accounting for the overall differences between NdNiO$_3$ and SmNiO$_3$, which have IMTs in very different temperature ranges.

Unfortunately, it is difficult to disentangle the contributions due to the resistivity change across the IMT and the substrate thermal conductivity. Both parameters decrease as temperature is increased. Within the nickelate family, it is observed that as the transition temperature increases (with the reduction of the rare-earth radius), the resistivity drop across the IMT decreases [2,6]. A similar trend is observed in the V$_2$O$_3$, VO$_2$, V$_3$O$_5$ family [10]. Therefore, it is not feasible to compare a material featuring a high temperature IMT with a large resistivity change, with another system having a low temperature IMT with a small resistivity change.

A way to overcome this drawback is to compare, for the same material, samples with different IMT quality. We fabricated two NdNiO$_3$ films: one with a high quality IMT, and a second one subjected to a 120°C, 30 minutes annealing in vacuum. The annealing creates oxygen vacancies, reducing the resistivity change across the IMT, as seen in Fig. 5a. This has clear consequences in the resistive switching properties, which are smoother for the annealed sample (Fig. 5b). Furthermore, there are notable differences in the filament, as can be seen in Figs. 5c and 5d. The annealed device shows much less contrast and homogeneity within the metallic area, perhaps due to the formation of multiple filaments. This differs from the well-defined, intense filament for the non-annealed sample, and points out to a smaller degree of metallization and current focusing. This confirms that the resistivity change across the IMT is a key parameter controlling resistive switching and filament characteristics, although does not rule out important contributions from thermal conductivity.

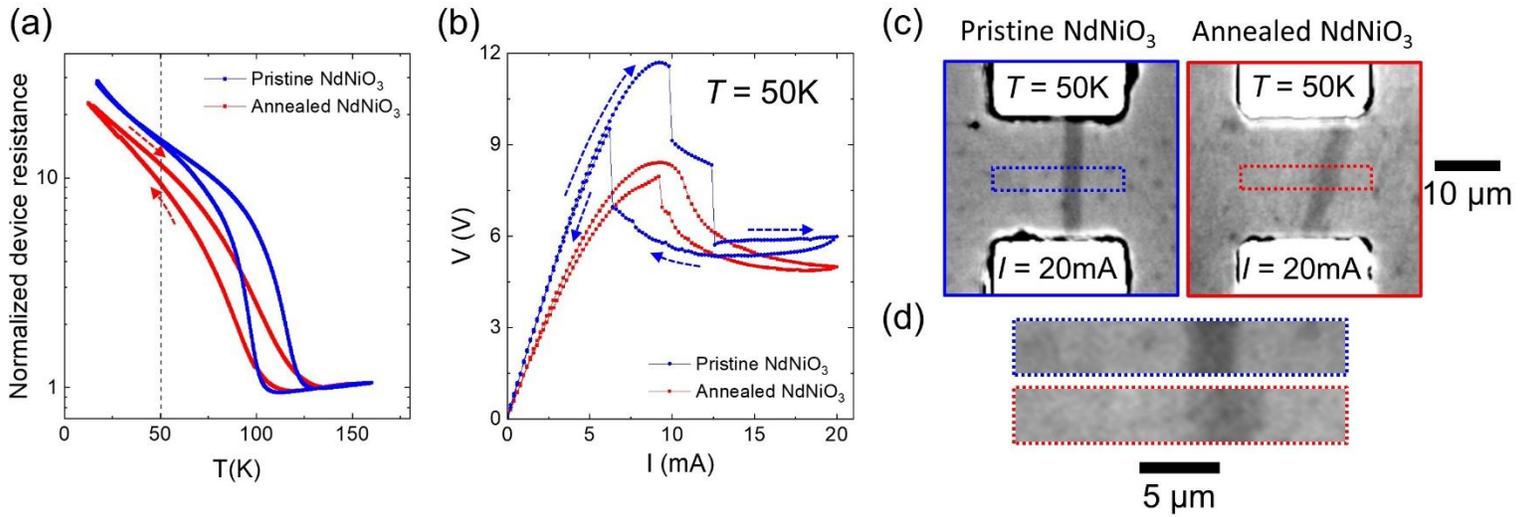

FIG. 5. Resistive switching and filament characteristic in pristine and annealed NdNiO$_3$. (a) Two-probe device resistance vs temperature on pristine (blue) and annealed (red) NdNiO$_3$ films. (b) Voltage vs current at $T = 50$ K for a pristine (blue) and annealed (red) sample. (c) Wide-field microscopy image of filament formation in pristine (left) and annealed (right) NdNiO$_3$. $T = 50$ K and $I = 20$ mA in both cases. Reflectivity was normalized using an area far from the gap region. (d) Zoomed image into the central part of the filament for pristine (top) and annealed (bottom) NdNiO$_3$. $T = 50$ K and $I = 20$ mA in both cases.

**Conclusions**

In summary, we have used a combination of *in-operando* standard and scanning near-field optical microscopies to study the characteristic lengths of filament formation during the electrically-induced IMT. We found that, in addition to bias current, filament width is strongly dependent on base temperature and the specific material. Lower set-up temperatures yield thinner filaments, increasing current density and local temperature, leading to sharper resistive switching properties. With the aid of resistor network simulations, we discussed the material properties that control filament size, underlining the importance of the resistivity drop across the IMT as well as the substrate's thermal conductivity.

Our results support recent works concerning another fundamental aspect of the electrically-induced IMT: switching dynamics [10,11]. It was proposed that a large resistivity ratio between insulator and metal would induce higher current focusing, increasing local Joule heating within the filament and explaining the different switching timescales observed in $V_2O_3$, $VO_2$, $V_3O_5$, $NdNiO_3$ and $SmNiO_3$. However, direct evidence of this has been lacking so far. The present work provides a systematic study of the characteristic lengthscales of the electrically-induced IMT, unveiling a strong connection between resistivity, thermal properties, filament size and resistive switching characteristics. The mechanisms outlined here are simple and general, and could be applicable to other types of resistive switching, such as ReRAM. Considered together with recent developments in the field [10,11,21], it completes a simple and unified picture of the length- and time-scales of filament nucleation, growth and relaxation, as well as underlining their importance for developing new technologies based on the IMT.

**Acknowledgements**

The authors thank Marco Lopes for his support during the fabrication and measurement of these samples. The sample fabrication and project coordination were funded by the Swiss National Science Foundation through an Ambizione Fellowship (#PZ00P2_185848). The oxide growth was supported by the European Research Council under the European Union's Seventh Framework Program (FP7/2007-2013)/ERC Grant Agreement 319286 Q-MAC and the Swiss National Science Foundation Project no. 200020-179155. W.R. was supported by the U.S. Office of Naval Research through the NICOP Grant N62909-21-1-2028. s-SNOM measurements were supported by the Swiss National Science Foundation through a Research Grant #200020_201096. Simulations were funded by the French ANR project "MoMA" ANR-19-CE30-0020. T. L. acknowledges support by the Cusanuswerk, Bischöfliche Studienförderung. J.d.V acknowledges support from the Spanish Ministry of Science through a Ramón y Cajal Fellowship (#RYC2021-030952-I).

Supplementary Information

# Characteristic lengthscales of the electrically-induced insulator-to-metal transition

Theodor Luibrand[†1], Adrien Bercher[†2], Rodolfo Rocco[†3], Farnaz Tahouni-Bonab[1], Lucia Varbaro[2], Carl Willem Rischau[2], Claribel Domínguez[2], Yixi Zhou[2], Weiwei Luo[2], Soumen Bag[3], Lorenzo Fratino[3,4], Reinhold Kleiner[1], Stefano Gariglio[2], Dieter Koelle[1], Jean-Marc Triscone[2], Marcelo J. Rozenberg[3], Alexey B. Kuzmenko[2], Stefan Guénon[1] and Javier del Valle[*2,5]

[1]Physikalisches Institut, Center for Quantum Science (CQ) and LISA+, Eberhard Karls Universität Tübingen, Auf der Morgenstelle 14, Tübingen 72076, Germany
[2]Department of Quantum Matter Physics, University of Geneva, 24 Quai Ernest-Ansermet, 1211 Geneva, Switzerland
[3]Laboratoire de Physique des Solides, UMR8502 CNRS - Université Paris-Sud, Université Paris-Saclay, 91405 Orsay Cedex, France
[4]Laboratoire de Physique Théorique et Modélisation, CNRS UMR 8089, CY Cergy Paris Université, 95302 Cergy-Pontoise Cedex, France
[5]Department of Physics, University of Oviedo, C/ Federico García Lorca 18, 33007 Oviedo, Spain

[†]These authors contributed equally to this work
*Corresponding author: javier.delvalle@unige.ch

**1. Methods**

Thin film and device fabrication

We grew $NdNiO_3$ and $SmNiO_3$ oxide films on (001) oriented $LaAlO_3$ substrates using off-axis magnetron sputtering in an $Ar:O_2$ (3.5:1) mixture at a pressure of 180 mTorr and substrate temperature of 460º C. Films are ~40-45 nm thick and grow epitaxially, as can be seen using X-ray diffraction (Figure S1c and S1d).

For microdevice fabrication, a combination of techniques was used. First, we patterned isolated $NdNiO_3$ and $SmNiO_3$ islands using optical lithography and Ar ion beam milling. This allows us to measure each device independently, since it is electrically isolated from the others. After this we patterned Pt electrodes on top of these islands. For that we used optical lithography followed by on-axis Pt sputtering at room temperature and a lift-off in acetone. Pt thickness is around 40 nm and the gap size is 20 μm x 20 μm. For the s-SNOM measurements, a further lithographic step was used. Optical lithography does not create smooth electrode edges. This is very challenging for SNOM, since the tip is tripped by the electrode irregularities. To improve this, we used electron beam lithography and a second Pt evaporation to define 10 μm x 10 μm electrodes with smooth edges.

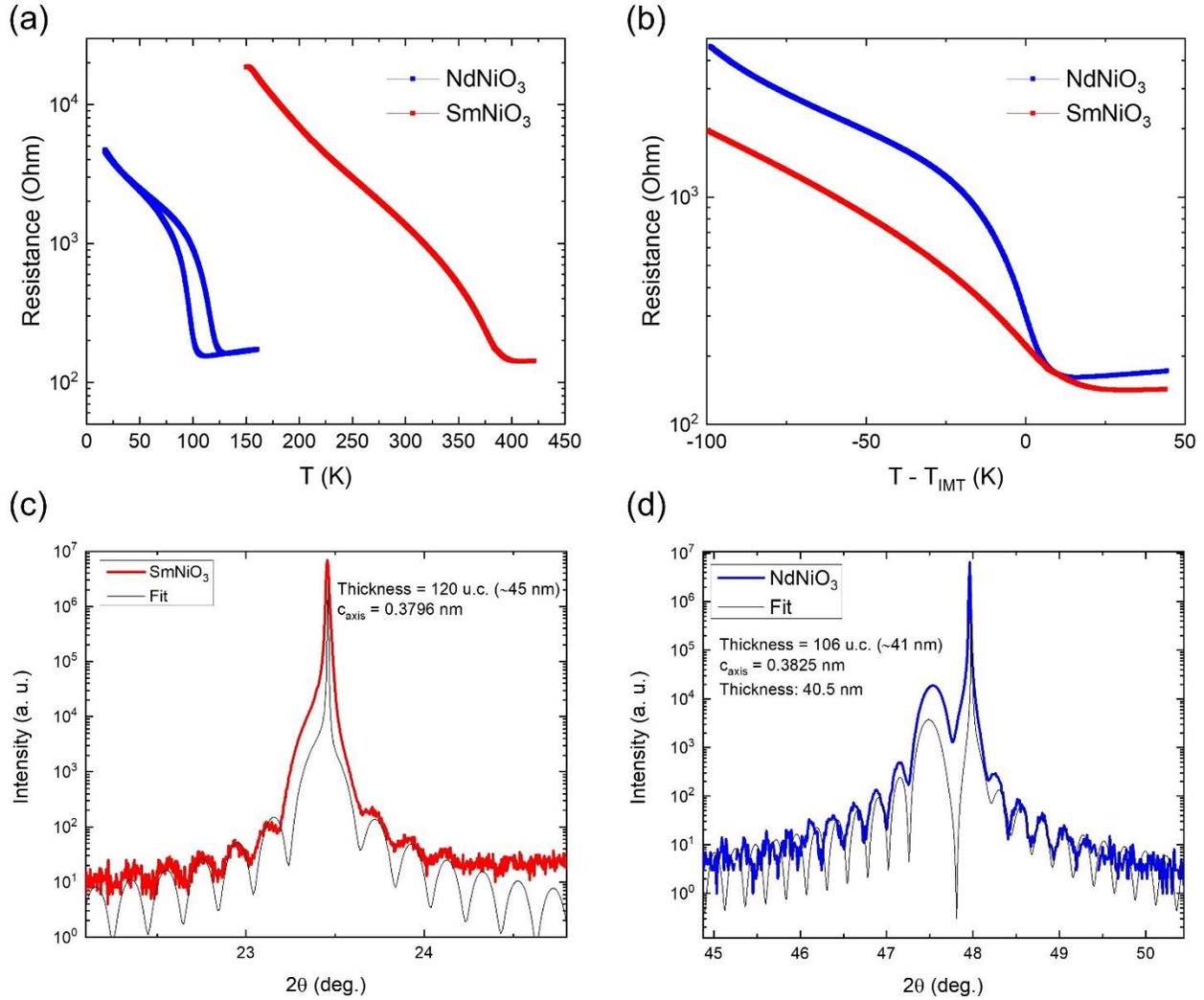

FIG. S1. Structural and transport properties of nickelate films and microdevices. (a) Two-probe resistance vs temperature characteristics of $NdNiO_3$ and $SmNiO_3$ microdevices, respectively. (b) Two-probe resistance plotted as a function of $T-T_{IMT}$. $T_{IMT}$ was calculated finding the maximum of $\partial \log(\rho)/\partial T$, were $\rho$ is the resistivity. Only the warm up branch is shown. $NdNiO_3$ shows a much sharper IMT. (c), (d) $\theta$-$2\theta$ scans of the $SmNiO_3$ and $NdNiO_3$ films, respectively. Scans were performed around the (001) $LaAlO_3$ peak. Finite size oscillations are very visible, allowing us to determine the film thickness. Data was manually fitted using the InteractiveXRDfit software [1], and is shown in the plots as continuous dark lines. XRD shows high quality epitaxial films with thicknesses around 40-45 nm.

*In operando* standard optical microscopy

We used an optical wide-field microscope that facilitates simultaneous imaging and electrical transport measurements [2]. The device under investigation is mounted in vacuum, on the cold finger of a liquid Helium continuous flow cryostat with a temperature range between 4.2 K and 300 K. The microscope has a spatial resolution of 500 nm, the illumination is a monochromatic

LED with a wavelength of 532 nm, and the maximum field of view is approximately 500 µm x 500 µm.

The electric transport properties were measured in a two-probe configuration. For NdNiO$_3$, we used a Keithley 2400 SourceMeter configured as a current source, whereas we used a highly stable self-built current source for SmNiO$_3$.

*Image Processing*: For NdNiO$_3$, the grey values were normalized to a NdNiO$_3$ area that is not influenced by the resistive switching (not in-between the electrodes). For SmNiO$_3$, all the images are differential – the image with zero bias current is subtractedd for each current.

## *In operando* s-SNOM

A cryogenic s-SNOM system (cryo-neaSNOM from neaspec/attocube GbmH) was used for nanoscopic imaging of the filaments in the devices. Infrared radiation from a Quantum Cascade Laser (Daylight Solutions) was focused at a metal-coated AFM tip (ARROW-NCPt-50 from NanoAndMore GmbH), which was grounded in order to reduce the electrostatic interaction between the tip and the sample. Despite that, we had to avoid applying voltages higher than 10 V in the microdevices as it would disturb the AFM in tapping mode. The tip size determines the spatial resolution (20 nm in this case). Pseudoheterodyne detection allows separating far-field and near-field contributions to the signal by using higher-order tapping harmonics (the 3rd harmonic is used in the present paper). The detected near field signal has an excellent spatial contrast between the insulating and metallic phases because of the large change of the optical conductivity across the IMT. More information about the s-SNOM operation can be found in [3].

## Resistor network simulations

In the simulations presented in this work we use a phenomenological mesoscopic model known as Mott Resistors Network [4, 5]. The model describes the material as a matrix of cells, each containing four resistors, which connect the cell to its four nearest neighbors. Each cell corresponds to a small region of the material of the order of 10 nm. This scale is chosen in order to define a phase for the cell, which can be insulating or metallic. At first, all the resistors are initialized to a high insulating value, and all the cells are in the insulator phase. A voltage is applied to the mesh through the metallic electrodes that are situated at the top and the bottom and currents begin to circulate in the resistor network. These currents can be computed, knowing the initial resistance and the applied voltage, using Kirchhoff laws. When the current $I$ flows through the resistors $R_{ij}$, these generate heat according to Joule's law, wit power $P = I^2 R$. The heat generated by a cell is given by the sum of the contributions of its four resistors

$$P_{ij}(t) = \left(I_1^2(t) + I_2^2(t) + I_3^2(t) + I_4^2(t)\right) R_{ij}(t)$$

where $t$ indicates time in units of the simulation time-step, $ij$ are the indexes which identify the cell, $P_{ij}$ is the power generated by the cell, $R_{ij}$ is the resistance of the four resistors (which are always assumed to share the same value) and $I_1, I_2, I_3, I_4$ are the four currents flowing through them. Setting to unity the geometrical dimension of the cell, we can identify the resistivity of the cell with $R_{ij}$. The temperature of the cell will be the result of two contributions: Joule heating and a dissipative term that includes the dissipation to the nearest neighbour cells and the dissipation to a substrate at a fixed temperature $T_0$, with which all the cells are in contact. Therefore, using the heat transfer equation, we can write the temperature of the cell as follows

$$T_{ij}(t) = T_{ij}(t-1) + \frac{P_{ij}(t)}{C} - \frac{K}{C}\left(5T_{ij}(t-1) - \sum_{kl}^{NN} T_{kl}(t-1) - T_0\right)$$

where $K$ is the thermal conductivity, $C$ the thermal capacity and the sum with indexes $kl$ runs over the nearest neighbour cells. We note that we have made the non-essential assumption of choosing the same thermal conductivity for the dissipation to the substrate and to the nearest neighbours, and that the time-step of the simulation $\Delta t$ is set to unity.

The first order transition of a cell from the insulator to the metal phase (and vice-versa) is described as a thermally activated behavior with a probability that depends on the temperature of the cell according to the following Arrhenius-like law

$$p_{ij}^{a\to b}(t) = exp\left(\frac{-E_B^{a\to b}(T_{ij})}{T_{ij}(t)}\right)$$

where $a$ and $b$ are the states of cell (insulator or metal) and $E_B$ is the energy barrier that separates the two corresponding local minima as described by the following Landau-type free energy (which is appropriate for a 1$^{st}$ order thermally driven transition)

$$f(\eta) = h\eta + p\eta^2 + c\eta^4$$

$$h = h_1 \frac{T - T_C}{T_C} + h_2$$

$$p = p_1 \frac{T - T_C}{T_C} + p_2$$

$\eta$ is the order parameter and $T_C$, $h_1$, $h_2$, $p_1$ and $p_2$ are constants. The resistivity of the cell is then chosen according to the state of the cell: low and constant ($\rho_{met}$) in the metal state and high and temperature dependent in the insulator state ($\rho_{ins}(T)$). In particular, we choose Mott's equation for variable range hopping [6] to describe the temperature dependence of the resistivity in the insulating state, since it has already been used to fit the resistivity of NdNiO3 samples [7, 8]

$$\rho_{ins}(T) = \rho_0 e^{\Delta\left(\frac{1}{T} - \frac{1}{T_{IMT}}\right)^{1/4}}$$

where $\Delta$ is a constant, $T_{IMT}$ is the metal-insulator transition temperature and $\rho_0 = \rho(T_{IMT})$ is the resistivity at the transition temperature. Nevertheless, the specific choice of the functional form does not change the main qualitative features of the results. Once the resistivity of the cell has been computed we can update the resistance of the resistors within it. When all the cells have been updated, the time-step is increased by one and the simulation continues as described above, starting again from the computation of Kirchhoff currents.

## 2. Filament size vs. Applied current

Apart from temperature and specific material choice (the focus of the paper), filament size depends strongly on the applied current. This is a well-known effect [9-13]. It can be best appreciated in the supplementary videos 1-10. Fig. S2 shows images of a NdNiO$_3$ microdevice at different points of a *V-I* measurement cycle, for a temperature $T = 60$ K. As can be seen, the filament width grows with applied current, appearing and disappearing at the discontinuities of the ramp-up and ramp-down *V-I* curves, respectively.

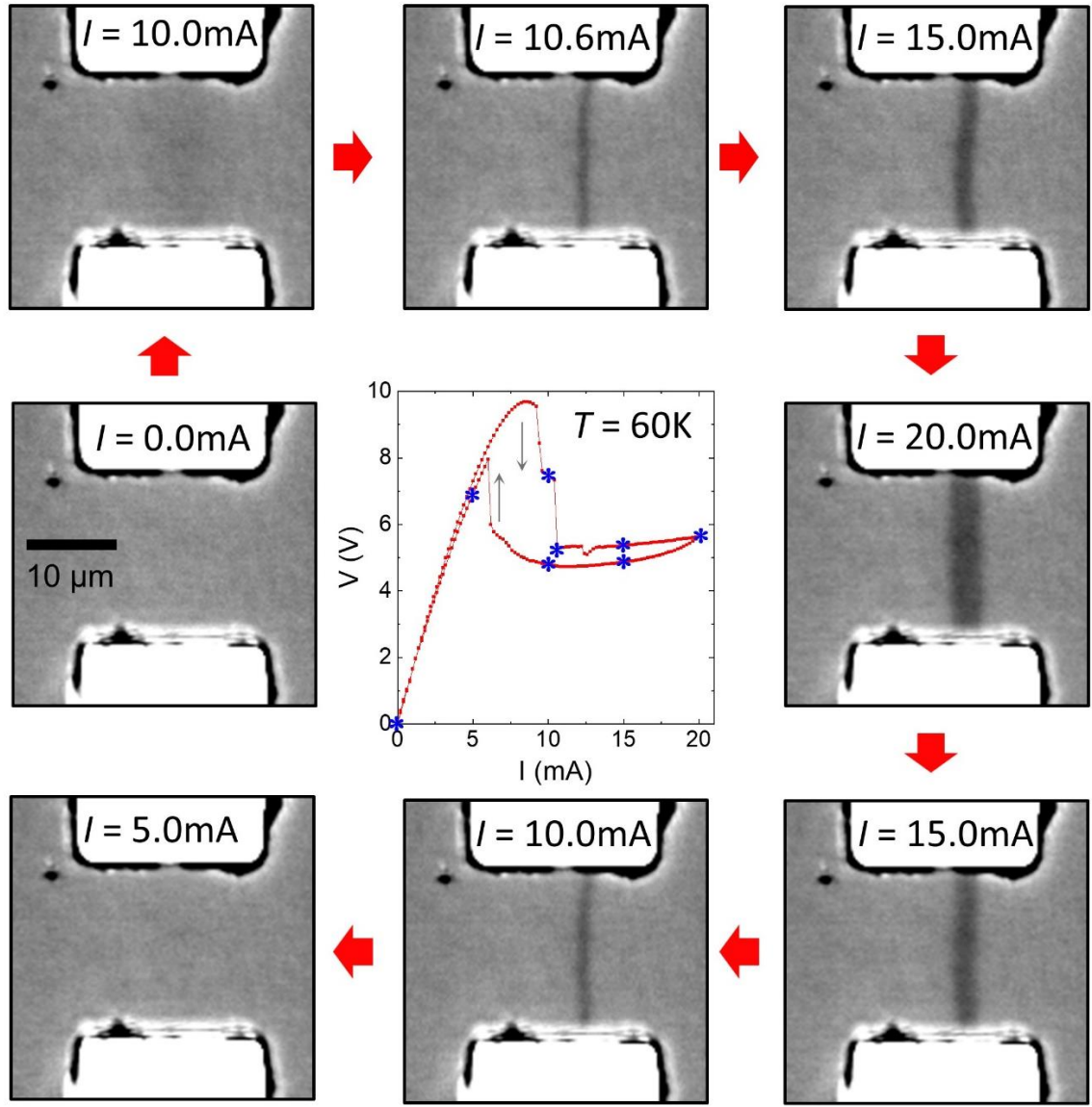

FIG. S2. Filament size dependence on applied current. Central panel: *V-I* characteristics of a NdNiO$_3$ microdevice at $T = 60$ K. Blue asterisks mark currents at which the outer panels where captured. Outer panels: standard optical microscope images of the devices at different applied currents.

## 3. Self-oscillation of microdevices

Interpretation of filament size is not straightforward for all currents. As can be seen in Fig. 2a and 2d in the main text, for some temperatures there are two voltage discontinuities when current is ramped up (for instance, at $T$ = 50 K in NdNiO$_3$). In between the two, there is a range of currents where the *V-I* curve is smooth. In that range, the system is not stationary, but rather oscillates between two configurations, one with a percolating filament and one without it. These self-oscillations, which are in the 10 kHz range, are a well-known effect [14,15] and they can be observed with an oscilloscope. The parasitic capacitance and the slow reaction time of current source are the main factors determining the oscillation regime.

All the analysis about filament size at different temperatures and materials shown in the paper is done for $I$ = 20 mA. This is well-above the self-oscillation range, where the system is stationary again, so it does not affect our conclusions.

**Supplementary references**